\documentclass[pra,twocolumn,showpacs,amsmath,amssymb,aps,a4paper,10pt]{revtex4-1}

 \usepackage{graphicx}
 \usepackage{dcolumn}
 \usepackage{bm}
 \usepackage{amssymb,amsmath}
 \usepackage{natbib}
 \usepackage{verbatim}
 \usepackage{subfigure}
 \usepackage{textcomp}
\usepackage{unicode}
\usepackage[utf8]{inputenc}
\usepackage[T1]{fontenc}
\usepackage{multirow}
\usepackage{booktabs}
\usepackage{ctable}
 \newcommand{\YSO}{Y$_2$SiO$_5$ }
\newcommand{\PRYSO}{Pr$^{3+}$:Y$_2$SiO$_5$ }
\newcommand{\otoprule }{\midrule [\heavyrulewidth ]}

\begin{document}

  \title{Coherent spectroscopy of rare-earth-ion doped whispering-gallery mode resonators}

 \author{D. L. McAuslan}
\author{D. Korystov}
 \author{J. J. Longdell}
 \email{jevon.longdell@otago.ac.nz}
 \affiliation{Jack Dodd Centre for Photonics and Ultra-Cold Atoms, Department of Physics, University of Otago, Dunedin, New Zealand.}
  \date{\today}
\begin{abstract}


We perform an investigation into the properties of \PRYSO whispering gallery mode resonators as a first step towards achieving the strong coupling regime of cavity QED with rare-earth-ion doped crystals. Direct measurement of cavity QED parameters are made using photon echoes, giving good agreement with theoretical predictions. By comparing the ions at the surface of the resonator to those in the center it is determined that the physical process of making the resonator does not negatively affect the properties of the ions. Coupling between the ions and resonator is analyzed through the observation of optical bistability and normal-mode splitting.

\end{abstract}

  \pacs{ 82.53.Kp, 42.50.Pq, 32.70.Cs, 42.65.Pc}

\maketitle

\section{Introduction}

The interaction between light and matter is of great importance in the field of quantum information science \cite{kimble08}. The introduction of an optical cavity into a light-matter system is an effective way of increasing the strength of the interaction \cite{Book-quantumoptics}. Of particular interest is the so called strong coupling regime of cavity quantum electrodynamics (QED) where the coupling between the atom and cavity is stronger than the decay rates of both the atom and the cavity \cite{kimble98}. In the strong coupling regime it is possible to detect and interact with single atoms \cite{mckeever03,boca04,trupke07}. The ability to address a single rare-earth-ion would be a marked development in the field of solid state quantum computing as it would fulfill the one remaining DiVincenzo criterion - a qubit-specific measurement capability \cite{divincenzo00,longdell04,wesenberg07}.

Rare-earth-ions have a number of properties that make them an interesting system for quantum information processing. They have very long optical \cite{sun02} and hyperfine coherence times \cite{fraval04, fraval05}. This gives rare-earth-ion doped crystals the ability to store photonic states for long periods of time, as demonstrated in \cite{fraval05, longdell05}. It has been shown that efficient quantum memories can be created using ensembles of ions \cite{hedges10}. Recently rare-earth-ions have been used to demonstrate creation of a high bandwidth memory \cite{saglamyurek11}, and storage of large numbers of qubits in a single memory \cite{usmani10, bonarota11}. Previous investigations into rare-earth-ions in optical cavities have looked at photon echoes for classical signal processing \cite{wang98}, non-linear effects due to strong-coupling between multiple atoms and a cavity \cite{ichimura06}, and the properties of ions and resonators required to achieve the strong coupling regime \cite{grudinin06,mcauslan09,goto10}.

When trying to reach the strong-coupling regime in rare-earth-ion systems, whispering gallery mode (WGM) resonators are an obvious choice due to their ability to simultaneously have the required large quality factors and small mode volumes. McAuslan et al. \cite{mcauslan09} have shown theoretically that the `bad cavity' strong coupling regime should be possible using millimetre sized resonators with quality factors $>10^{9}$. Here the properties of \PRYSO WGM resonators are measured as a first step towards performing strong coupling cavity QED experiments. 

We measure the coherence time, population lifetime, and spectral hole lifetime of the ions near the surface of the resonator. These are compared to the properties of Pr$^{3+}$ ions in the center of the resonator, and in a bulk sample. If the process of making the resonator has damaged the crystal structure a degradation of these properties for ions near the surface should result. We measure the atom-cavity coupling strength and compare it to the theoretical value based on the transition dipole moment of the ions and the resonator mode volume. Atom-cavity interactions are observed through optical bistability and normal-mode splitting and compared to a theoretical model. We finish off with a brief discussion of the effect near field radiative heating has on coherence time measurements. Throughout the paper we demonstrate the distinct advantage of this system - the ability to make direct measurements of cavity QED parameters utilizing simple photon echo techniques.

\section{Experimental Setup}

Two \YSO WGM resonators were used in this work with Pr$^{3+}$ concentrations of 0.005\% (resonator A) and 0.05\% (resonator B). The resonators are fabricated from single \PRYSO crystals using mechanical grinding and polishing techniques. The crystal is oriented such that the vertical axis of the resonator corresponds to the b-axis of the crystal, i.e. light is coupled into the D1-D2 plane. The resonators are spheroidal in shape and both have a radius of 1.95~mm. The quality factor of the resonators used in this experiment range from $1-4 \times10^6$ (different resonator modes have slightly different quality factors).

\begin{figure}[t]
  \centering
  \includegraphics[width=0.48\textwidth]{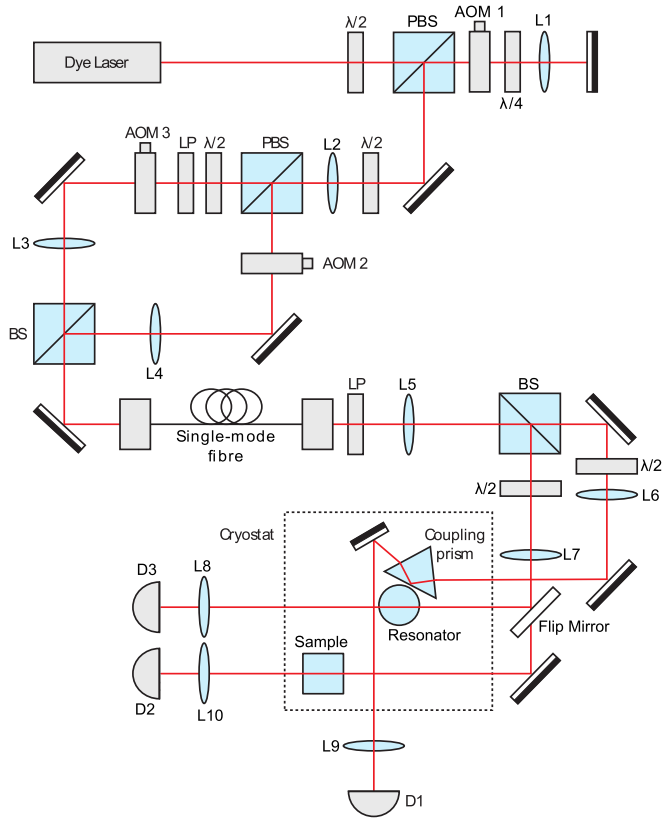}
\caption{\label{fig:expsetup}(Color online)
 Experimental Setup. PBS - Polarizing beamsplitter, BS - non-polarizing beamsplitter, LP - linear polarizer, AOM - acousto-optic modulator. The resonator, coupling prism and sample are all mounted inside a cryostat and cooled to 3.2~K. AOM1 is used for scanning and ranges from 70-90~MHz, AOM2 is at +80~MHz, AOM3 is at +125~MHz creating a 45~MHz beat between the signal and local oscillator paths. AOM2 is used to make the echo pulses. L1, L8, L9, L10=50~mm, L5, L6, L7=100~mm, L2,L3,L4=200~mm. A flip mirror is used to alternate between having the second beam go through the resonator or the bulk sample.
 }
\end{figure}

The experimental setup is shown in Fig. \ref{fig:expsetup}. A Coherent 699 dye laser stabilized to less than 5~kHz drives the $^3$H$_4$ - $^1$D$_2$ transition in \PRYSO at 605.977~nm. A Mach-Zehnder interferometer is used to facilitate heterodyne detection of the echoes. The interferometer is set up such that both the probe beam and the local oscillator (LO) beam pass through the coupling prism, which enables better mode matching between the two beams. Two acousto-optic modulators (AOM's) are used to shift the frequency of the LO by 45~MHz with respect to the probe. 45~MHz has been chosen because it is greater than the total hyperfine splittings in \PRYSO (36.92~MHz \cite{equall95}), which means the LO does not interact with any of the ions being driven by the probe. AOM2 gates the laser to create the pulses used in the echo sequence. AOM3 gates the laser so that the LO is only turned on when an echo is expected. A single mode fibre is used to clean the mode of the laser beam, increasing coupling into the resonator.

\begin{figure}[t]
  \centering
  \includegraphics[width=0.48\textwidth]{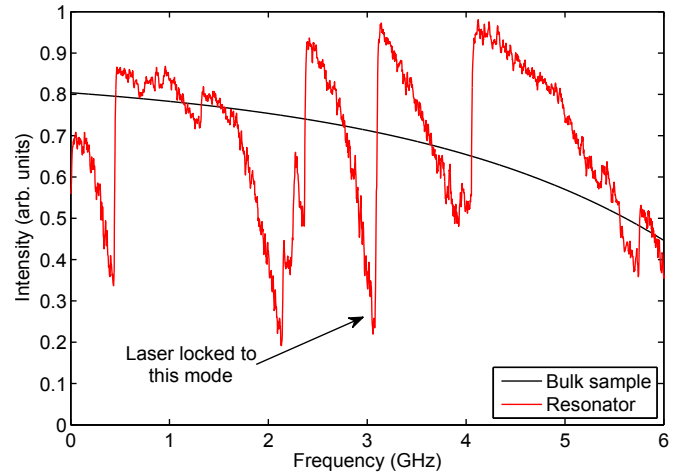}
\caption{\label{fig:location}(Color online)
 Location of the resonator modes with respect to the inhomogeneous line of the Pr$^{3+}$ ions. The black line corresponds to transmission through the \PRYSO bulk sample. The red line corresponds to the laser intensity after the coupling prism. Dips in intensity correspond to coupling into resonator modes. Note the assymetrical shape of the resonator modes due to optical bistability.}
\end{figure}

The resonator, coupling prism and sample are all mounted inside a vibration-isolated, closed-cycle cryocooler (manufactured by S2 corporation) and cooled to 3.2~K. Light is evanescently coupled into the resonator using a cubic zirconia prism ($n_c = 2.174$). An Attocube Systems ANPx101 linear positioner allows adjustment of the distance between the resonator and coupling prism at cryogenic temperatures. The ability to alter this distance is critical to enable efficient coupling into the resonator. The polarization of the light is rotated to maximize the coupling (up to 30\%). A 5x5x5~mm 0.02\% \PRYSO crystal is used as a reference to compare the location of the resonator modes to the inhomogeneous line of the ions (see Fig. \ref{fig:location}). 

\begin{figure}[t]
  \centering
  \includegraphics[width=0.48\textwidth]{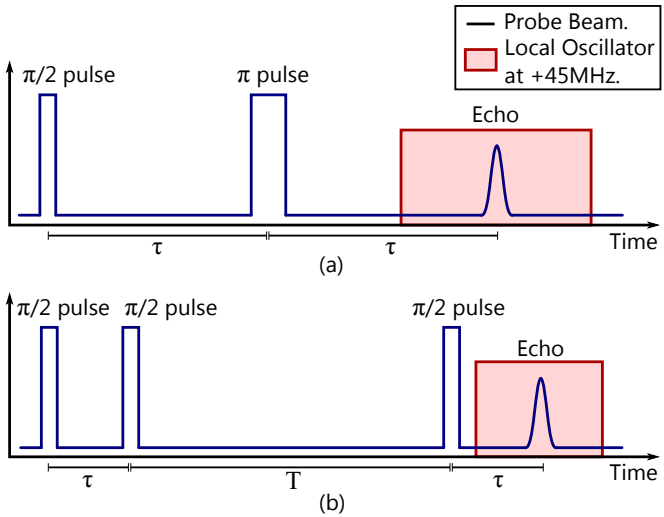}
\caption{\label{fig:echo}(Color online)
Pulse sequences involved in a (a) 2PE, (b) 3PE. In these experiments the echo is detected using heterodyne detection with the local oscillator shifted 45~MHz from the probe beam.
}
\end{figure}

The majority of the experiments in this work use either two pulse photon echoes (2PE) or three pulse photon echoes (3PE) to measure properties of the ions. The sequence of pulses used to create echoes is shown in Fig. \ref{fig:echo}. Figure \ref{fig:echo}(a) shows the standard two pulse photon echo sequence. In this work the two pulse photon echo is used to make two sets of measurements; the atom-cavity coupling ($g$), and the coherence time ($T_2$) of the ions in the resonator. Figure \ref{fig:echo}(b) shows the sequence of pulses that is used to create a 3PE. Here 3PEs are used to measure $T_1$ in the resonator.

\section{Results}

\begin{figure}[t]
  \centering
  \includegraphics[width=0.48\textwidth]{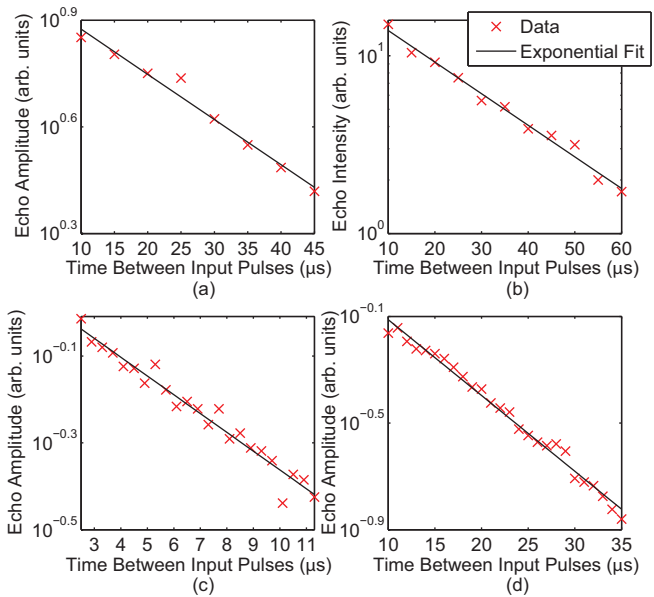}
\caption{\label{fig:resonatorecho}(Color online) The size of the echo formed as the time between the two input pulses is varied with (a) light coupled into resonator A, (b) the laser through the bulk sample, (c) light coupled into resonator B, (d) the laser propagating through resonator B. From the exponential fits the coherence times are measured to be (a) 68~$\mu$s, (b) 98~$\mu$s, (c) 21.0~$\mu$s, (d) 30.8$~\mu$s.
}
\end{figure}

By measuring how the size of the echo changes as the time between the $\pi/2$ and $\pi$ pulses is altered the coherence time can be determined. This is shown in Figs. \ref{fig:resonatorecho}(a) and (b) for light coupling into resonator A and for a beam passing through the bulk sample.

Echoes in the resonator were measured using heterodyne detection (which measures the amplitude), therefore to calculate the coherence time $e^{-2 \tau/T_2}$ is fit to the data \cite{Book-spectroscofsolidsrareearth}. This gives the coherence time in resonator A to be $T_2 = 68~\mu$s. Echoes in the bulk sample were measured using direct detection because they were large enough that heterodyne detection was not required. Direct detection measures the intensity of the field, therefore $e^{-4 \tau/T_2}$ is fit to the bulk sample data \cite{Book-spectroscofsolidsrareearth}. The coherence time in the sample is measured to be $T_2 = 98~\mu$s. Modelling the echo decay as being exponential provides a good fit to the data.

To measure the effect that instantaneous diffusion (ISD) has on the coherence time of the ions further echo experiments were performed using resonator B (see Figs. \ref{fig:resonatorecho}(c) and (d)). $T_2$ was measured for the laser coupled into the resonator, and with a beam passing through the resonator, with input powers ranging from 0.2 - 1.8~mW in steps of 0.1~mW. The length of the input pulses were altered so that the pulse areas were always $\pi/2$ and $\pi$. The coherence time in resonator B was measured to be $T_2 = 21.0~\mu$s, and with the beam going through the resonator $T_2 = 30.8~\mu$s. Over the range of input powers the coherence time was found to vary by $\pm3~\mu$s. The variation appears random so is attributed to experimental error in the measurement process and not ISD. 

We note that $T_2$ in resonator B is significantly less than that measured in the resonator A. This is because it was necessary to alter the experimental setup inside the cryostat. We believe this caused the resonator to be warmer than in previous measurements.

\begin{figure}[t]
  \centering
  \includegraphics[width=0.48\textwidth]{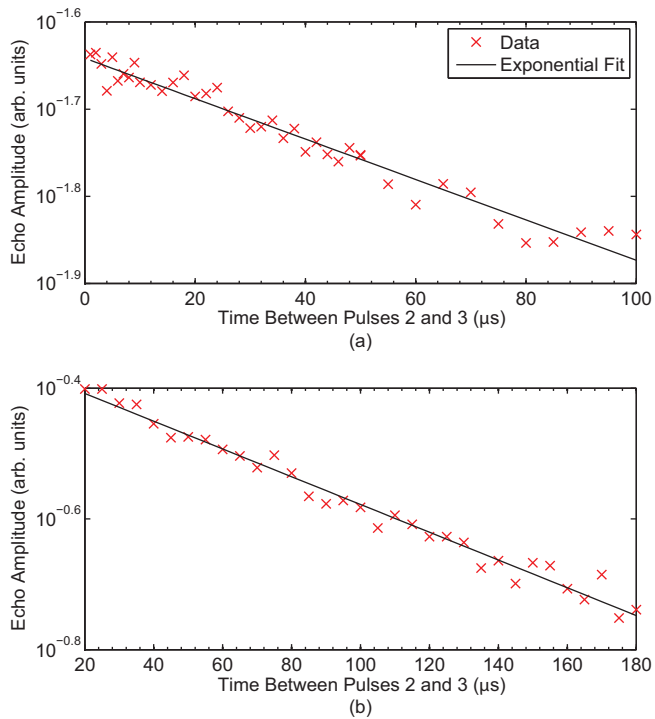}
\caption{\label{fig:T1measure}(Color online) The size of the 3PE formed as the time between pulse 2 and 3 ($\mathrm{T}$) is varied with (a) light coupled into the resonator and (b) the laser propagating through the resonator. By fitting $e^{-\mathrm{T}/T_1}$ to the data the population lifetimes are measured to be (a) $T_1 = 187~\mu$s, (b) $T_1 = 205~\mu$s.
 }
\end{figure}

Three pulse photon echoes were used to measure the population lifetime of the ions in resonator B. By measuring how the size of the echo decays as the time between pulses 2 and 3 is increased, $T_1$ can be measured (see Fig. \ref{fig:T1measure}). $e^{-\mathrm{T}/T_1}$  is fit to the data and the population lifetimes are measured to be $T_1 = 187~\mu$s (coupled into the resonator) and $T_1 = 205~\mu$s (through the resonator). These two measurements are in relatively good agreement, and also agree well with that measured by Equall et al. \cite{equall95} ($T_1 = 164~\mu$s) by observing the fluorescence decay in a 0.02\% \PRYSO crystal.

The spectral hole lifetime was measured using accumulated three pulse photon echoes in resonator B. This technique involves using optical pumping to create a frequency grating in the atomic population that builds up due to population storage in another hyperfine level \cite{hesselink79, hesselink81, shelby83, saikan89}. Pulses of light can be stored in this grating, and by varying the time between preparation and storage the hole lifetime can be measured. 

\begin{figure}[t]
  \centering
  \includegraphics[width=0.48\textwidth]{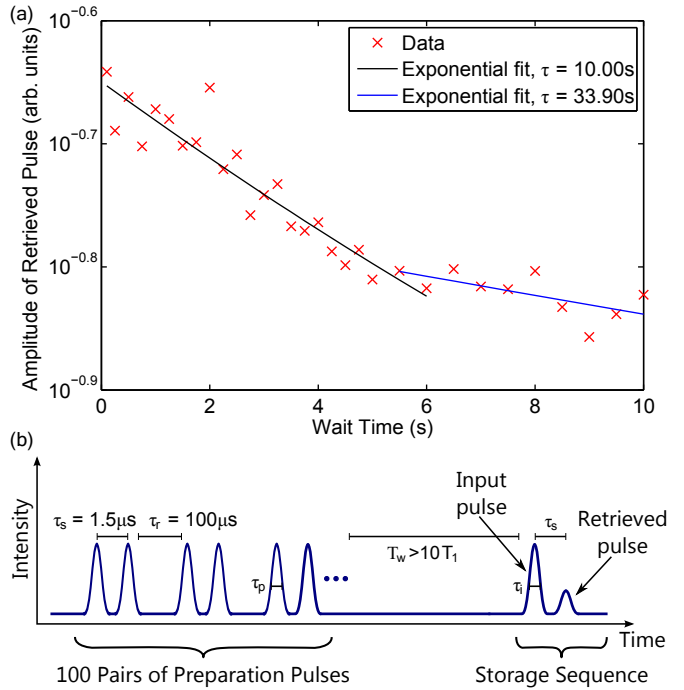}
\caption{\label{fig:holelifetime} (Color online) (a) Measurement of the lifetime of spectral holes in a \PRYSO WGM resonator using accumulated photon echoes. (b) Pulse sequence used in the experiment. $\tau_s$ - separation between preparation pulses, $\tau_r$ - time between pairs of preparation pulses, $\mathrm{T}_w$ - wait time between grating preparation and pulse storage/retrieval, $\tau_p$ - length of preparation pulses, $\tau_i$ - length of input pulse.
 }
\end{figure}

The pulse sequence used in this experiment is shown in Fig. \ref{fig:holelifetime}(b). 100 pairs of pulses ($\tau_p =100$~ns) separated by $\tau_r > T_2$ optically pump the ions to create the grating. After the grating is prepared there is a waiting time, $\mathrm{T}_w > 10T_1$, that ensures the excited state is empty. Pulses of light are applied to the sample and stored for $\tau_s$. Over the time $\mathrm{T}_w$ the grating will decay due to relaxation of the spin lattice \cite{shelby80, holliday93}, and the rate of this decay will be determined by the spectral hole lifetime. As the grating decays the storage efficiency will decrease. By measuring the size of the retrieved pulse as $\mathrm{T}_w$ is increased the hole lifetime can be determined (see Fig. \ref{fig:holelifetime}(a)).

From Fig. \ref{fig:holelifetime}(a) there appears to be two rates of decay in this system, measured by fitting $e^{-\mathrm{T}_w/T_h}$ to the data. There is an initial decay with a decay constant of 10.0~s, followed by a secondary decay with decay constant of 33.9~s. These hole lifetimes are of the same order of magnitude measured previously in bulk \PRYSO samples \cite{holliday93, fraval05-thesis}. The lifetime of spectral holes has a large dependence on temperature. Based on this we infer the temperature of the resonator is $\sim6$~K \cite{holliday93, ham99}.

\begin{figure}[t]
  \centering
  \includegraphics[width=0.48\textwidth]{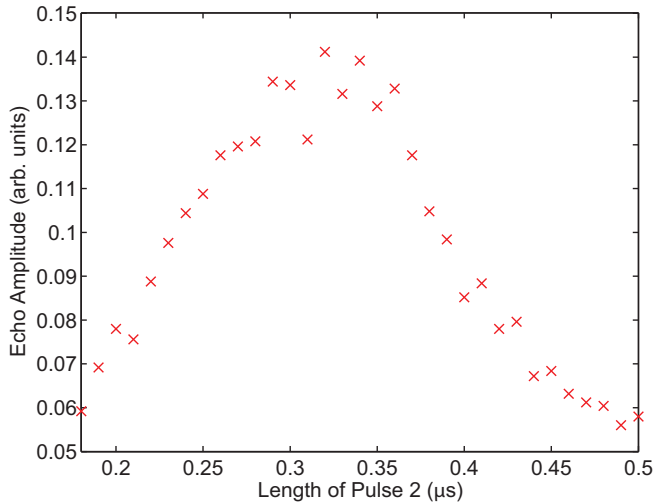}
\caption{\label{fig:echoarea}(Color online) Amplitude of the echo as the area of the second input pulse is varied for resonator B. The maximum echo amplitude should correspond to the input pulses having areas of $\pi/2$ and $\pi$ respectively.
 }
\end{figure}

Using two pulse photon echoes the size of a $\pi$ pulse is determined, thus giving a measure of the atom-cavity coupling strength. The size of the echo is measured as the length of the second pulse is altered at fixed power (see Fig. \ref{fig:echoarea}). Assuming the maximum echo amplitude corresponds to the area of the second pulse being equal to $\pi$, the length of a $\pi$ pulse is $0.32~\mu$s. The Rabi frequency of the ions ($\Omega$) is calculated from the area ($\Theta$) and length ($\tau$) of the input pulses:

\begin{align}
 \Omega & =  \frac{\mu E}{\hbar} \nonumber \\
	& =  \frac{\Theta}{\tau}
\end{align}

giving $\Omega = 9.82$~MHz. The coupling between a single photon and an atom in the resonator is calculated by dividing the Rabi frequency by the number of intracavity photons ($n_{\text{ph}}$):

\begin{equation} \label{eq:coupling}
g = \frac{\Omega}{2 \sqrt{n_{\text{ph}}}}
\end{equation}

For an input power of $700~\mu$W and 20.6\% coupling into a resonator with a quality factor of $ 1.8 \times 10^{6}$, the number of intracavity photons is $1.28 \times 10^{5}$. This corresponds to an atom-cavity coupling of $g= 2\pi \times 1.73$~kHz.

A summary of the properties of the resonators measured in this work is displayed in table \ref{tab:resonatorparams}.

\begin{center}
\ctable[ mincapwidth = .5\textwidth,
 caption = {\label{tab:resonatorparams} Properties of 0.005\% (A) and 0.05\% (B) \PRYSO WGM resonators. 2PEs were used to measure both $T_2$ and $g$. $g$ was also measured from an analysis of optical bistability. $T_1$ was measured using 3PEs, and the hole lifetime was measured using accumulated 3PEs. All values have either been compared to a beam going through the resonator/bulk sample ($T_1$, $T_2$), theoretical calculations ($g$), or values stated in the literature (hole lifetime).} , botcap]{ccc}
{
\tnote{Measured in bulk sample.}
\tnote[b] {Measured with beam going through resonator.}
\tnote[c] {Measured using photon echoes.}
\tnote[d] {Calculated from optical bistability modeling.}
\tnote[e] {Calculated from theoretical mode volume.}
\tnote[f] {From ref. \cite{ham99}.}
}{
\multicolumn{3}{c}{ \textbf{Resonator A}} \\
\vspace{-3mm} \\
\FL
 & Resonator& Comparison \\
\otoprule
$T_2 (\mu$s) & 68 & 98\tmark[a]\LL
\vspace{-0mm} \\
\multicolumn{3}{c}{ \textbf{Resonator B}} \\
\vspace{-3mm} \\
\FL
& Resonator& Comparison\\
\otoprule
$T_1 (\mu$s)  &187 & 205\tmark[b]\\
$T_2 (\mu$s) & 21.0& 30.8\tmark[b]\\
$g/2\pi$ (kHz) & 1.73\tmark[c],  2.2\tmark[d]& 2.47\tmark[e] \\
Hole Lifetime (s) & 10.0, 33.9 & 50-400\tmark[f]\LL
}
\end{center}

\section{Discussion of Photon Echo Experiments}

The first set of experiments involved making measurements on the ions close to the surface of the resonator to determine whether the process of creating the resonator has affected the properties of these ions. Of particular concern is whether $T_2$ has been affected as any decrease in coherence time will make it more difficult to achieve strong coupling \cite{mcauslan09}. Here the optical coherence time, population lifetime, and hole lifetime of the ions were measured. When measuring $T_1$ and $T_2$  two sets of measurements are made, one with the laser coupled into a resonator mode and another with a beam passing either through the bulk sample or the resonator. Because WGMs are confined close to the surface of the resonator the coupled beam should interact only with the surface ions and thus measure their properties. The majority of the ions interacting with the beam passing through the resonator will not be near the surface. This gives a measure of the properties of the rest of the ions (which would not be affected by making the resonator), which the surface ions can then be compared to.

We were concerned that instantaneous spectral diffusion (ISD) could be causing a difference in coherence time between the resonator and bulk sample. ISD is an effect where using intense pulses in the echo sequence decreases the measured coherence time \cite{taylor74,huang89,kroll90,equall95}. The laser field inside the resonator is amplified due to the high quality factor, therefore the ions would experience a larger field than for a beam going through the sample. Thus it was expected ISD would cause a reduction in $T_2$ for the resonator. However the coherence time measurements performed in resonator B showed no evidence of the existence of ISD.

The coherence time in resonator A is $31\%$ smaller than that in the bulk sample, and for light coupled into resonator B it is $32\%$ smaller than for a beam going through the resonator. If making the resonator had damaged the crystal structure at the resonator surface a large reduction in coherence time would be expected. One explanation for this difference in $T_2$ is that the temperature of the resonator surface could be different to the center of the resonator/bulk sample. The only thermal contact the resonator has with the cryostat cold head is through a 1.6mm diameter aluminium post that it is mounted on. To gain an understanding of what is required to raise the temperature of the resonator, we calculate that a heat load of 0.22~mW would create a 1~K difference in temperature between the surface and center of the resonator. It is feasible that heat loads of this magnitude could be caused by residue gases in the cryostat. It has been shown previously that the measured $T_2$ can have a significant dependence on the sample temperature \cite{konz03}, as at increased temperatures phonon scattering becomes the dominant dephasing mechanism \cite{equall95}.

The requirement of the `bad cavity' strong coupling regime is that $\frac{g^2}{\kappa} > \gamma$ ($\kappa$, $\gamma$ are the cavity and atomic decay rates) \cite{mcauslan09}. Therefore large atom-cavity coupling is desirable. We compare the measured value of $g$ ($=2\pi \times 1.73$~kHz) to the theoretical value to ensure it is as large as expected. The theoretical value of $g$ can be calculated from the transition dipole moment of the atoms ($\mu$) and the resonator mode volume ($V$). $V$ is calculated by integrating over the field inside the resonator:

\begin{equation}
 V = \frac{\int \epsilon(r) |E|^2 d^3r}{\mbox{max}(\epsilon(r) |E|^2)}
\end{equation}

The mode volume of the fundamental mode of a 1.95~mm spherical resonator is $5.40 \times 10^{-13}$~m$^3$. The atom-cavity coupling is:

\begin{equation}  \label{eq:coupling2}
g = \frac{\mu}{n_r} \sqrt{\frac{\omega_a}{2 \hbar \epsilon_0 V}}
\end{equation}

where $\omega_a$ is the transition frequency of the ions, $n_r$ is the refractive index of the resonator. From equation (\ref{eq:coupling2}), $g=2\pi \times 2.47$~kHz. There is good agreement between the calculated and measured values of atom-cavity coupling.

\section{Cavity QED parameters}

One method of defining the strong coupling regime is by using the saturation photon number ($n_0$) - the number of photons required to saturate the transition of an intracavity atom, and the critical atom number ($N_0$) - the number of atoms necessary to affect the field inside the cavity. The strong coupling regime can be defined as $(N_0, n_0)<1$ \cite{mcauslan09}. The critical atom number and saturation photon number can be related to the atom and cavity parameters by the following equations:

\begin{equation}\label{eq:n_0}
N_0 \equiv \frac{2 \gamma_h \kappa}{g^2} \quad \mbox{and} \quad n_0 \equiv \frac{\gamma \gamma_h}{4g^2}
\end{equation}

where $g$ is defined in equations (\ref{eq:coupling}) \& (\ref{eq:coupling2}), $\kappa = \frac{\pi c}{\lambda Q}$, $\gamma = \frac{1}{T_1}$, and $\gamma_h = \frac{1}{T_2}$.

Cavity QED parameters are calculated for the resonator used in these experiments. For resonator A $g= 2 \pi \times 1.73$~kHz, $\kappa = 2 \pi \times 138$~MHz, $\gamma_h = 2 \pi \times 2.34$~kHz, and $\gamma = 2 \pi \times 851$~Hz. This gives $N_0 = 2.15 \times 10^{5}$, $n_0 = 0.166$. This shows that for rare-earth-ion doped cavities the requirement that $n_0 < 1$ is easily achieved even with millimeter sized resonators. On the other hand $N_0$ is 5 orders of magnitude too large, but improvements in this area should be possible. It has been shown that it is possible to achieve quality factors up to $5.3 \times 10^{10}$ in crystalline resonators \cite{grudinin06}.

McAuslan et al. \cite{mcauslan09} performed an investigation into different materials which could be used in rare-earth-ion doped cavity QED and showed that there are available materials better suited for cavity QED experiments than \PRYSO. For example a Er$^{3+}$:\YSO resonator with radius = 1~mm and $Q=1 \times 10^{10}$ would give critical numbers of $n_0 = 1.88 \times 10^{-5}$ and $N_0 = 0.104 $, which is well into the strong coupling regime.

\section{Optical Bistability and Normal Mode Splitting}

When the coupling between the atoms and cavity are strong enough such that the presence of the atoms has an appreciable affect on the cavity mode optical bistability and normal mode splitting result. These effects are well understood and have been studied extensively over the last several decades \cite{rempe91, rosenberger91, thompson92, brecha95, gripp97, ichimura06}. These phenomena in rare-earth-ion doped resonators have been studied previously by Ichimura and Goto \cite{ichimura06}. Here we follow their methods to describe observations in our resonators.

 As a simple model of the holeburning dynamics that occur in the resonator the ions are treated as being two-level systems all with the same transition frequency ($\omega_a$), fixed at some frequency higher than that of the laser ($\omega_c$). Using the stationary solution to the Maxwell-Bloch equations given in \cite{ichimura06}:

\begin{equation} \label{eq:bistable}
y = x \left[ \left( 1 + \frac{g^2 N \chi}{\gamma_h \kappa} \right) + i \left( \frac{\omega_c - \omega_l}{\kappa} - \frac{g^2 N (\omega_a - \omega_l) \chi}{\gamma_h^2 \kappa} \right) \right]
\end{equation}

where $|y|^2 \ $ is the input field, $|x|^2$ is the output field, and $N$ is the number of atoms interacting with the cavity mode. $\chi$ is the susceptibility, approximated by using the form of a plane-wave ring cavity \cite{rosenberger91}:

\begin{equation} \label{eq:chi}
\chi = \frac{1}{2 |x|^2} \text{ln} \left(1 + \frac{2 |x|^2}{1 + \frac{(\omega_a-\omega_l)^2}{\gamma_2^2}} \right)
\end{equation}

\begin{figure}[t]
  \centering
  \includegraphics[width=0.48\textwidth]{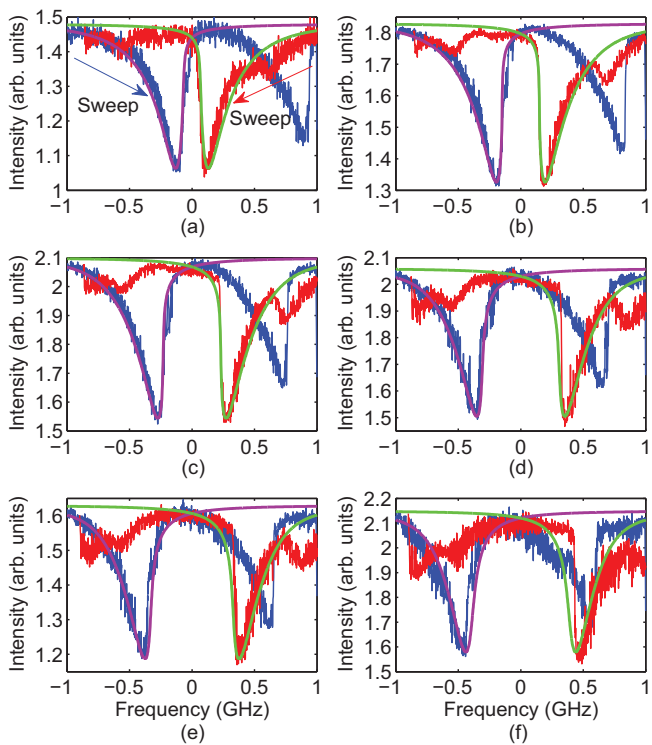}
\caption{\label{fig:bistability}(Color online) Observation of optical bistability and normal mode splitting in resonator B for varying input powers. (a) 800~$\mu$W, (b) 400~$\mu$W, (c) 200~$\mu$W, (d) 100~$\mu$W, (e) 80~$\mu$W, (f) 40~$\mu$W. The blue plots correspond to the laser being swept in the forward direction, i.e. from -1 to 1~GHz.  The red plots correspond to the laser being swept in the reverse direction, i.e. from 1 to -1~GHz. The purple and green lines correspond to the cavity mode calculated from equation (\ref{eq:bistable}).
 }
\end{figure}

Figure \ref{fig:bistability} shows how the cavity mode is affected by the presence of the atoms for input powers ranging from $800~\mu$W - $40~\mu$W. Optical bistability and normal mode splitting are clearly visible. As the laser is swept in the forward direction one normal mode peak is resolved, the other becoming apparent as the laser is swept in the opposite direction.

By fitting equation (\ref{eq:bistable}) to the data in Fig. \ref{fig:bistability} the atom-cavity coupling is estimated to be $2\pi \times 2.2$~kHz. This agrees well with the values measured using photon echoes ($2\pi \times 1.73$~kHz), and calculated from the resonator mode volume ($2\pi \times 2.47$~kHz). The parameters used in this model are $\kappa = 2\pi \times 123$~MHz, $\gamma_h = 2\pi \times 7.58$~kHz, $\gamma = 2\pi \times 2.34$~kHz, $N = 1.6 \times 10^8$. The coupling into the resonator mode is 28.7\% and the loss due to reflection from mirrors and cryostat windows is estimated to be 0.2.

\section{Near Field Radiative Heating}

\begin{figure}[t]
  \centering
  \includegraphics[width=0.48\textwidth]{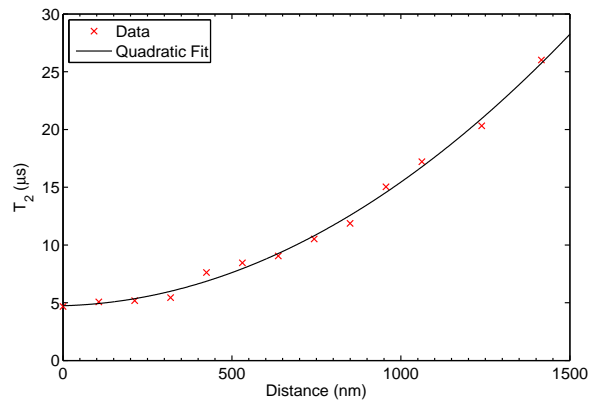}
\caption{\label{fig:stepsvsT2}(Color online) Variation in coherence time as the distance between the prism and resonator is varied. The rate of heat transfer exhibits a $\frac{1}{d^2}$ dependence \cite{joulain05}, therefore a quadratic has been fit to the data. The distance axis is calibrated from how many steps the ANPx101 positioner has moved running in open-loop mode at 15~V. At 4~K a 15~V step corresponds to a displacement of $\sim$35~nm, but as this movement is temperature dependent the distance between prism and resonator is only approximate.
 }
\end{figure}

Whilst performing these experiments we noticed the measured coherence time appeared to be dependent on the distance between the resonator and the coupling prism (see Fig. \ref{fig:stepsvsT2}). This behavior was observed both for light coupled into the resonator and for a beam passing through it. We explain this strange behavior as being caused by near field radiative heating of the resonator. 

In initial experiments there was limited thermal contact between the coupling prism and cryostat cold finger. When the distance between the prism and resonator was small (not in contact) thermal radiation emitted from the prism caused the resonator to heat up. This phenomena has been studied by Joulain et al. \cite{joulain05} who show that for distances less than $1~\mu$m thermal radiation causes a significant transfer of heat between two objects with different temperatures. To verify that the variation in coherence time was indeed caused by radiative heating the experimental setup was modified to ensure good thermal contact between the prism and cold finger. This removed the dependence of $T_2$ on coupling distance.

\section{Conclusion}

We have measured the properties of a WGM resonator doped with rare-earth-ions. The atom-cavity coupling has been measured using two pulse photon echoes, and also by modeling optical bistability/ normal mode splitting in the resonator. The measured values of $g$ were shown to be in good agreement with the theoretical value calculated from the resonator mode volume and atom transition dipole moment. The coherence time, population lifetime and hole lifetime of the atoms at the resonator surface are measured and compared to those in the center of the resonator. We infer from these measurements that the process of making the resonator does not have a large effect on the properties of the ions. An analysis of cavity QED parameters is performed and it is determined that the strong-coupling regime should be attainable by making resonators with larger quality factors and smaller mode volumes. Near field radiative heating of the resonator by the coupling prism is investigated and shown to significantly affect the temperature of the resonator.

\section{Acknowledgments}

DLM, DK and JJL were supported by the New Zealand Foundataion for Research Science and Technology under Contract No.\ NERF-UOOX0703.

\end{document}